# Electrochemical Impedance spectroscopy study of AgI-Ag$_2$O-MoO$_3$ Glasses: Understanding the Diffusion, Relaxation, Fragility and Power Law Behavior

B. Tanujit, S. Asokan*

Department of Instrumentation and Applied Physics, Indian Institute of Science, Bangalore – 560 012, India

**Abstract**

Electrochemical Impedance Spectroscopy and Raman studies are performed on fast ion conducting, AgI-Ag$_2$O-MoO$_3$ glasses, over a wide range of composition to understand the features of structure, ion migration and their correlation. These features essentially involve diffusion and relaxation. The coefficients associated with diffusion process, especially, the diffusion coefficient, diffusion length and relaxation time has been determined by applying Nguyen-Breitkopf method. Besides, by tuning the concentration of the constituents, it is possible to obtain samples those exhibit two important structural characteristics: Fragility and Polymeric phase formation. The present study essentially addresses these issues and endeavors to figure out the corroboration among them. The relaxation behavior, when scrutinized in the light of Diffusion Controlled Relaxation model, ascertains the fragility threshold which is also identified as the margin between the two types of polymeric phases. Simultaneously, it fathoms into the equivalent circuitry, its elements and their behavioral changes with above mentioned features. The power law behavior of A.C. conductivity exhibits three different non-Jonscher type dispersive regimes along with a high frequency plateau. The sub-linearity and super-linearity remain significantly below and above the Jonscher's carrier transport limit, $0.5 \leq n \leq 0.9$. Finally, by observing the behavior of the crossover between these sun-linear and super-linear (SLPL) regime, an intuitive suggestion has been proposed for the appearance of SLPL: oxygen vacancy formation at higher frequency.

___

* Corresponding Author email: sundarrajan.asokan@gmail.com, sasokan@iisc.ac.in

Fax: 91-80-23608686; Phone. +91-80-22933195, 91-80-22932271

## 1. Introduction

AgI based alkali molybdate glasses are one of the rich materials from many scientific and technical aspects. Colloquially, this solid material possesses a structure of super cooled liquid [1], exhibit fast ion conduction which is around $10^{-5}$-$10^{-2}$ $\Omega^{-1}$ $cm^{-1}$ at room temperature ($T_R$) [2]. For last few decades this material, as an affluent member of the class of fast ion conducting (FIC) glasses has been under thorough research to scientifically understand its structural basics, differences with other FIC glasses, physical properties [3-11] and technologically exploit it for diverse applications [12, 13]. But scientific endeavor is a never ending process because each exploration opens up many new layers of insights. Basically, the structure of this type of glasses involves the high temperature (~150$^O$C), fast ion ($Ag^+$) conducting α phase of AgI and a surrounding glass matrix, $Ag_2O$-$MoO_3$ in the present case, where this phase is dispersed. The role of this matrix is to hinder the phase transition of AgI at $T_R$ [14]. On the other hand, the ion migration involves Frenkel defect, oxygen vacancies in the form of non-bridging oxygen (NBO) and hopping [15]. Apparently, these aspects of structure and migration process implicate compelling electrochemical behaviors and properties. Thus to understand the correlation between the structure and migration, diffusion and relaxation, in the light of recent theoretical, experimental and analytical developments is one of the few challenges.

The ratio of glass matrix constituents i.e. $Ag_2O$/$MoO_3$, is a key factor to tune the NBO distribution over the matrix. This structural tuning can alter between two phases: over NBO formation instigates depolymerization of the matrix; conversely, their elimination or filling up results in polymeric phase formation. Besides, a simultaneous modulation in AgI mole percentage should not only enforce change in migration but also in structural fragility of the glass. Thus understanding how this fragility and polymeric phases incorporate with the electrochemistry and migration process appears to have emergent importance. Moreover, a coextensive challenge lies in explaining equivalent circuitry and a connection between the circuit elements with structure/migration correlation.

In the present study, these challenges are encountered for bulk AgI-$Ag_2O$-$MoO_3$ glasses, over a wide range of compositions, to accommodate diverse structural changes. The structural properties are studied employing Raman spectroscopy. The electrical and electrochemical properties are studied with impedance spectroscopy (EIS), by applying inter electrodes, to restrict intercalation and probe into the self-diffusion mechanism of the glass. Computational methods are employed to recognize the circuit elements and the whole circuitry. Furthermore, to explain the obtained results and establish the correlations among them, recent advanced theoretical analysis is applied. For instances, Nguyen-Breitkopf method (NB method) [16] for calculating diffusion coefficients from impedance data, Diffusion

Controlled Relaxation (DCR) model [17, 18] to explain relaxation behavior, Power law behavior to interpret A.C. conductivity ($\sigma_{ac}$) etc. Equipped with all these theoretical, experimental and numerical technique, this study revolves around identifying the fragility threshold, equivalent circuit, diffusion, relaxation and dispersive behavior of $\sigma_{ac}$.

## 2. Experimental
### A. Glass preparation

The essential and primary section of glass preparation by using microwave melt-quenching technique, its novelty and the reason for the selection of this sample domain i.e. the composition range, has been discussed in detail elsewhere [19]. Besides, the presence of three phased rigidity percolation (Floppy phase (FP), Intermediate phase (IP) and Stressed rigid phase (SRP)) in the strong glass forming region and nanocluster formation in the fragile region [20] has been recognized as a new concern while selecting the range of composition for the present work. Hence, 13 different samples, within a range of 53.75 ≤ x ≤ 39.5, almost randomly distributed among the four regions (Nano-cluster forming region, FP, IP and SRP) for bulk $(AgI)_x$-$(Ag_2O)_{25}$-$(MoO_3)_{75-x}$ solid electrolyte glasses has been prepared and quenched at room temperature ($T_{quenching} = 300\ K$) by pouring the melt into a stainless steel container of radius 5.12 ± 0.02 mm to make pallets of equal sizes. Samples in SRP and beyond, some of those are presented in this work, exhibit great difficulty in preparation. The excess configurational entropy ($\Delta S_c > 0$) due to increased covalent bond density per atom in these region, eventually results in high brittleness [21]. To avoid this inconvenience, a slow quenching method i.e. quenching at high temperature ($T_{quenching} > T_g$ of that particular sample) has been conducted for some of these samples, resulting better robustness. For EIS spectroscopy experiment, all the palletized samples are polished down to a thickness of 2.82 ± 0.02 mm.

### B. Material Characterizations

The EIS spectroscopy study of the palletized samples has been carried out with an Agilent 4294A Electrochemical Impedance analyzer, within a frequency range from 40 Hz to 10 MHz and a test signal level of 100 mV, without any DC bias. The 4 terminals of the instrument connect directly to the Guarded/Guard and Unguarded electrodes of a Keysight 16451B dielectric test fixture. Samples with appropriate dimension have been placed between the electrode after careful short and open calibration of the fixture. Raman study has been performed on solid samples, in a Horiba Jobin Yvon spectrometer. To avoid laser heating induced crystallization and fluorescence background, a near infrared source of 784.8 nm diode laser has been used. These spectra are taken in the frequency region from 50-1000 cm$^{-1}$ with spectral resolution for the Stokes side Raman of 0.7 cm$^{-1}$. The experimental data analysis has been done by using EIS Spectrum analyzer software, Python programming, Parallel computation in a Linux based platform and OriginPro.

## 3. Results
### A. Raman Spectroscopy

The obtained Raman data are normalized, base line corrected and reduced chi-square ($\chi_R$) fitted to a superposition of relevant Voigt profiles to get the mode centroid (ν) and mode-normalized scattering strength (I) values, using OriginPro software. The $\chi_R$ value remains within $10^{-4}$-$10^{-5}$. The Voigt profiles have been used because the present sample under investigation is glass with super cooled liquid structure, synonymously an amorphous solid. Below 200 $cm^{-1}$, a $Ag^+$ restrahlen mode [22] appear at 75 $cm^{-1}$; another peak at 105 $cm^{-1}$ represents $Ag^+$ ion vibration in the tetrahedral surroundings formed by $I^-$ ions [4]. Above 200 $cm^{-1}$, four peaks appear at 311 $cm^{-1}$, 335 $cm^{-1}$, 839 $cm^{-1}$ and 878 $cm^{-1}$ for samples within the composition criterion $Ag_2O/MoO_3 \geq 1$. These four peaks represent $\nu_2(E)$, $\nu_4(F_2)$, $\nu_3(F_2)$ and $\nu_1(A_1)$ modes of $MoO_4^{2-}$ monomeric tetrahedra [4, 8]. As the mole percentage ratio decreases i.e. $Ag_2O/MoO_3 < 1$; two more peaks appear at 781 $cm^{-1}$ and 895 $cm^{-1}$, along with the four $MoO_4^{2-}$ tetrahedral Raman modes. These modes have been argued [8] to be originated due to the formation of $[Mo_2O_7]^{2-}$ as in a structural form of infinite chain of corner sharing $MoO_6$ octahedra and bridging $MoO_4$ tetrahedra; similar to that of $Na_2Mo_2O_7$. The reaction involved in this polymeric phase formation is described as,

$$MoO_3 + MoO_4^{2-} \rightarrow [Mo_2O_7]^{2-} \tag{1}$$

### B. Impedance Spectroscopy

The quality and consistency of the impedance data obtained from the electrochemical impedance analyzer, for all the samples has been checked for Kramer-Kronig compliance test (KK test) in the EIS spectrum analyzer software. The KK test is based on the idea of causality and expressed as a correlation between the imaginary and real part of the response function in a physical system which is conditioned to be linear, time invariant, stable and finite within the whole frequency domain (0, ∞) [23]. The linear KK test in the present software is based on the model suggested by B. Boukamp [23]; it plots residuals between real/imaginary part of experimentally obtained data and that of a Voigt-RC (Parallel resistor-capacitor circuit units in series) model. The residuals remain around 0.1% and get slightly pronounced (~ 0.4%) for the imaginary part at higher frequency range; the time constant/data point ratio is kept at 0.5. This confirms a good quality experimental data and hence allowing further data processing and designing an equivalent Voigt-RC circuit model [23, 24]. **Figure 1(A)** and **1(B)** exhibit the Bode (|Z| (impedance) and φ (phase) vs. log $f$ (frequency)) and Nyquist (-Im|Z| (imaginary part of impedance) vs. Re|Z| (real part

of impedance)) plots for a representative sample, respectively to understand all the polarization processes involved and to model an equivalent circuit.

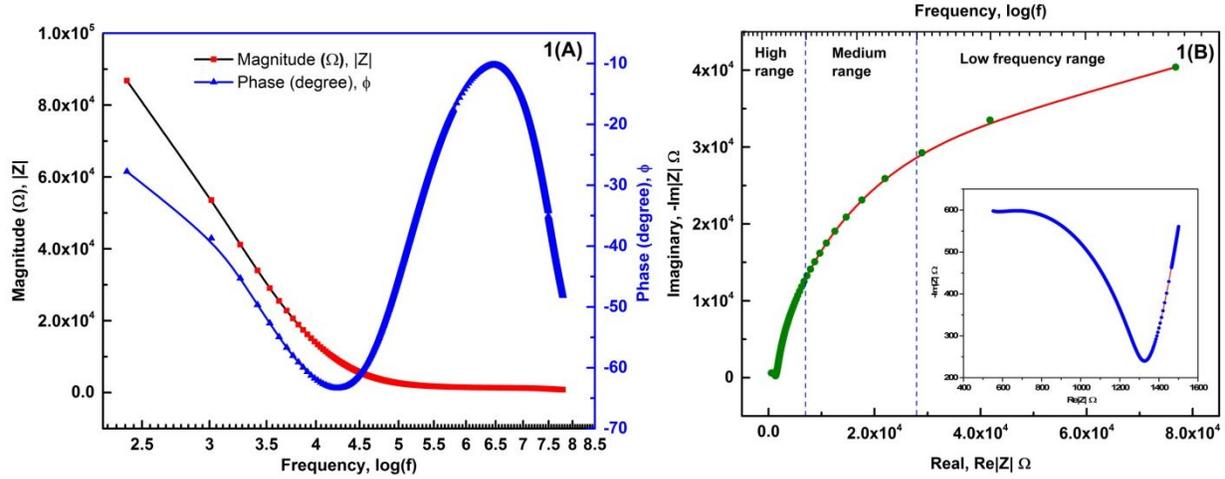

*Figure 1: (A) Bode plot; Magnitude (Ω) i.e. modulus of complex impedance Z, |Z| and phase (degree), φ vs. frequency (log f). (B) Nyquist plot; imaginary part of impedance, -Im|Z| vs. real part of impedance, Re|Z| for the whole frequency range, 40 Hz to 10 MHz (1(B)-Inset) -Im|Z| vs. Re|Z| plot in the high frequency range for a representative sample $(AgI)_{39.5}$-$(Ag_2O)_{25}$-$(MoO_3)_{35.5}$.*

### C. The Equivalent Circuit

These two, Bode and Nyquist plots over the entire frequency range (40 Hz to 10 MHz) together assist to distinguish the loss mechanisms involved during the electrochemical process. The plot of φ against log f in **Figure 1(A)**, exhibits two near capacitive behaviors near -10° and -65° for high and medium frequency range, respectively. These reflect upon the nature of Nyquist plot (**Figure 1(B)**) with two depressed semicircles. This observation is designed into two Voigt type RC circuits with different time constants, in series, represented as ($R_2$-$CPE_1$) and ($R_3$-$CPE_2$). Because of the depression in both the semicircles, the capacitors in these circuits is replaced by the constant phase elements (CPE) with impedance $Z(CPE_k) = (j\omega C_k)^{-n_k}$ where j is imaginary unit, ω is angular frequency, C is the capacitance in Farad and $n_k$ is an independent parameter with value $0 \leq n_k \leq 1$. At low frequency, a slight down bending in the φ vs. log f plot has been interpreted as the presence of diffusion process or rather interfacial phenomena which introduces a semi infinite Warburg element (Zw), in series with the two R-CPE circuits. This low frequency feature gets masked by relatively higher value of the second semicircle radius, in the medium frequency range and the limit of the present EIS instrument that performs unwell below 40 Hz. The Magnitude (Ω), |Z| plot (**Figure 1(A)**) resembles to that of a circuit with resistance and capacitance in series. The transition from capacitive to resistive behavior at log f ~ 4.5 is high enough to conclude that the capacitance has a lower value. This concludes an addition of a resistor ($R_1$) and a CPE ($CPE_3$), a replacement for the capacitor, in series with earlier designed circuit element. Furthermore, an inductor ($L_0$) is added in series that remains inactive below 100 kHz [24]. **Figure 2** shows the equivalent circuit model obtained from these observations and conclusions.

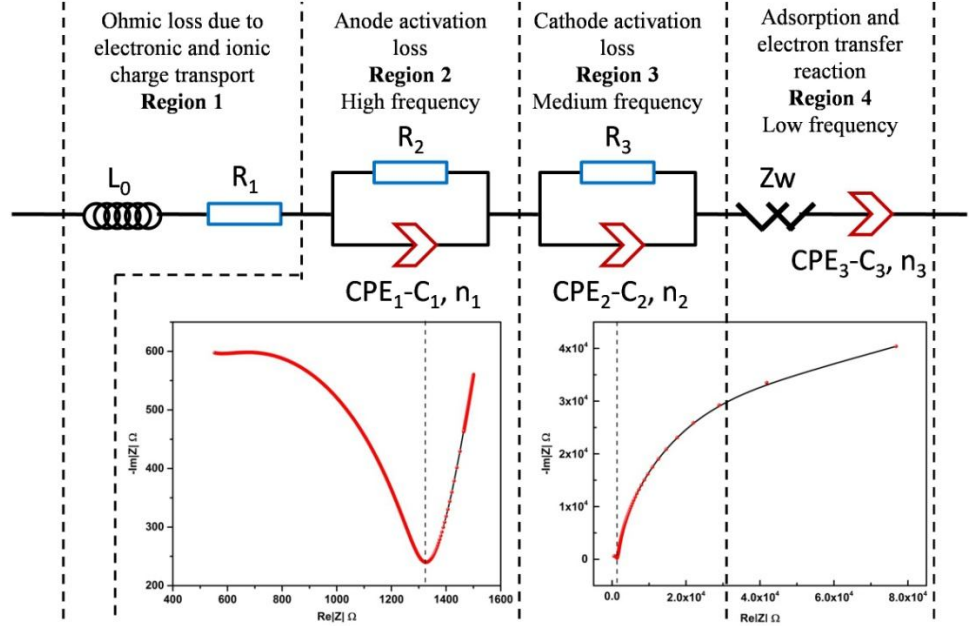

*Figure 2: Equivalent circuit model with a combination of Voigt-type analog, for bulk AgI-Ag$_2$O-MoO$_3$ glass of radius 10.24 ± 0.02 mm and thickness 2.82 ± 0.02 mm with inert electrodes. All the loss processes and adsorption process are showed in four different regions with corresponding Nyquist plot*

The impedances of these elements are;

$$ReZ(R_1) = R_1; \quad ImZ(R_1) = 0 \tag{2}$$

$$ReZ(L) = 0; \quad ImZ(L) = \omega L \tag{3}$$

$$ReZ(R_A - CPE_B) = \frac{R_A[1+(\omega C_B R_A)^{n_B} \cos(n_B \pi/2)]}{1+2(\omega C_B R_A)^{n_B} \cos(n_B \pi/2)+(\omega C_B R_A)^{2n_B}} \tag{4}$$

$$ImZ(R_A - CPE_B) = \frac{R_A[1+(\omega C_B R_A)^{n_B} \sin(n_B \pi/2)]}{1+2(\omega C_B R_A)^{n_B} \cos(n_B \pi/2)+(\omega C_B R_A)^{2n_B}} \tag{5}$$

$$ReZ(A_w) = \frac{A_w}{\sqrt{\omega}}; \quad ImZ(A_w) = \frac{A_w}{\sqrt{\omega}} \tag{6}$$

$$\text{ReZ(CPE}_3) = \frac{\cos(n_3\pi/2)}{(\omega C_3)^{n_3}}; \quad \text{ImZ(CPE}_3) = \frac{\sin(n_3\pi/2)}{(\omega C_3)^{n_3}} \qquad (7)$$

Thus the total impedance of the equivalent circuit is given as,

$$\text{ReZ(Eq)} = \text{ReZ}(R_1) + \text{ReZ}(L) + \text{ReZ}(R_2 - \text{CPE}_1) + \text{ReZ}(R_3 - \text{CPE}_2) + \text{ReZ}(A_w) + \text{ReZ(CPE}_3) \qquad (8)$$

$$-\text{ImZ(Eq)} = \text{ImZ}(R_1) + \text{ImZ}(L) + \text{ImZ}(R_2 - \text{CPE}_1) + \text{ImZ}(R_3 - \text{CPE}_2) + \text{ImZ}(A_w) + \text{ImZ(CPE}_3) \qquad (9)$$

### D. Physical Significances of Equivalent Circuit elements

The resistor, $R_1$ corresponds to the ohmic loss due to electronic and ionic charge transport and the inductor, $L_0$ has been introduced to balance the cell and cable inductivity. Now, the arrangement of the whole experiment is that a bulk fast ion ($Ag^+$) conducting glass of certain radius and thickness placed in between two inert electrodes those work like blocking electrodes. Hence, the electrode-electrolyte interface establishes a high work function that restrains any charge transfer other than electronic. This hindrance of ionic mass transport results in polarization near the interface. This process has been modeled as two Voigt type RC analogs in series ($R_2$-$CPE_1$) and ($R_3$-$CPE_2$) [25], suggesting anodic and cathodic activation loss processes at medium and high frequency range respectively. The replacement of capacitors of these two circuit elements with CPEs is interpreted as the imperfection of capacitors which is caused by the surface roughness that has a fractal like structure [26, 27] and that result in inhomogeneous charge distribution over the surface [28]. The parameter $n_k$ in the impedance equation of CPE i.e. $Z(CPE_k) = (j\omega C_k)^{-n_k}$ is a qualitative measure for the surface roughness; the more its value recedes from unity, the less smooth the surface becomes. This is evident because of using emery sheets for the purpose of smoothen the surface to make equal thicknesses for all the samples. The semi infinite Warburg element is comprised of contributions from three different processes (i) electronic diffusion ($W_{ion}$) and (ii) adsorption of ionic species on the surface of electrode-electrolyte interface ($W_{ads}$) because in terms of electroanalytical response, intercalation and adsorption processes are similar [29-31] (iii) a self-diffusion process instigated by ionic concentration gradient ($W_{diff}$). All these processes are explained in the discussion section. Surface roughness, loosely bonded surface atoms and defects instigate this adsorption process. Thus, this low frequency adsorption process along with the electron transfer reaction is represented as Frumkin and Melik-Gaikazyan impedance [31] which is a semi infinite Warburg in series with an adsorption type capacitor, replaced by CPE to consider the fractal roughness.

### E. Computational Fitting

Finally, after gathering all the physically meaningful circuit elements and thus designing the equivalent circuit, the experimental data for imaginary and real value of impedance are fitted, using a Chi-square test program written in Python to obtain the values of circuit elements. The program is parameterized with 11 parameters (P = 11), a set of ($R_1$, $L_0$, $n_1$, $n_2$, $n_3$, $R_2$, $R_3$, $C_1$, $C_2$, $C_3$, $A_w$) where $Z(CPE_k) = (j\omega C_k)^{-n_k}$; k = 1, 2, 3. The Chi-square test basically optimizes both the quantities

$$X^2(Re) = \sum_{i=1}^{k} \frac{(ReZ(Measured)_i - ReZ(Calculated)_i)^2}{ReZ(Calculated)} \qquad (10)$$

$$X^2(Im) = \sum_{i=1}^{k} \frac{(ImZ(Measured)_i - ImZ(Calculated)_i)^2}{ImZ(Calculated)} \qquad (11)$$

Where (k-1) are the degrees of freedom. These computations for 13 samples with k~1600 and P = 11 become very expensive and time consuming if done in a serial manner. The parallelization has been done as much efficiently as possible and tested by using 6 processors for each samples, in a Linux based platform. Afterwards, the calculated values for all 11 parameters for 13 samples, exhibit $X^2$ values much less than the $X^2$ value for 1600 degrees of freedom with 0.05 probability level and hence providing a good fitting. Figure 3 shows the Nyquist plot, fitted with the experimentally obtained impedance data for representative samples.

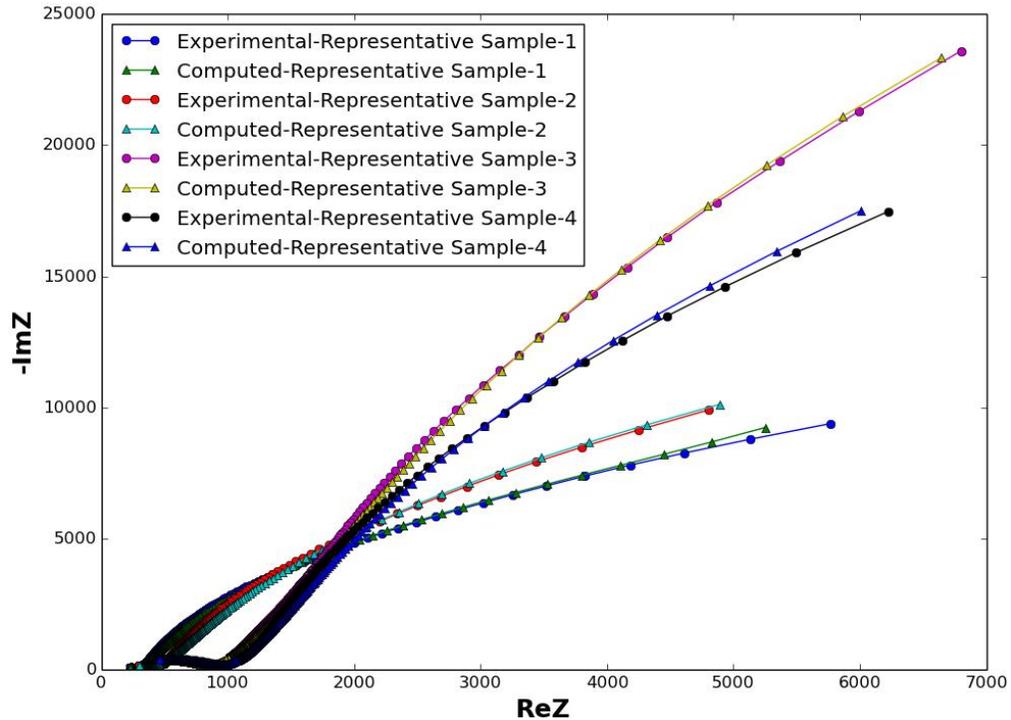

*Figure 3*: *Nyquist plot (within the whole frequency range; 40 Hz to 10 MHz) fitting by using Chi-square test for representative sample-1 (($AgI$)$_{52.5}$-($Ag_2O$)$_{25}$-($MoO_3$)$_{22.5}$), sample-2 (($AgI$)$_{50}$-($Ag_2O$)$_{25}$-($MoO_3$)$_{25}$), sample-3 (($AgI$)$_{46.25}$-($Ag_2O$)$_{25}$-($MoO_3$)$_{28.75}$) and sample-4 (($AgI$)$_{42.75}$-($Ag_2O$)$_{25}$-($MoO_3$)$_{32.25}$)*

## 4. Discussions

### A. Calculations of coefficients related to Diffusion

To ascertain a correlation between the ion diffusion and microstructure, as well as the underlying physical characteristics of the bulk AgI-Ag$_2$O-MoO$_3$ glasses, diffusion coefficient of net mobility ($D_m$), relaxation time distribution ($\tau$) in the double layer thickness which equals to Debye length ($\lambda_D$) and diffusion length ($l_D$) has been calculated, using Impedance spectroscopy data, by a method proposed by T. Q. Nguyen and C. Breitkopf [16] and corroborated with the structural features obtained from Raman spectroscopy. The method suggests as follows:

The diffusion coefficient of net mobility $D_m$, is defined as the ratio $\lambda_D^2$ and $\tau$,

$$D_m = \frac{\lambda_D^2}{\tau} \qquad (12)$$

In the effective permittivity model, the loss tangent ($\tan(\varphi)$) of an electrochemical system with two blocking electrodes, is given by the ratio of dielectric loss and real permittivity,

$$\tan(\varphi) = \frac{\text{Im}\varepsilon}{\text{Re}\varepsilon} = \frac{\text{Re}Z}{\text{Im}Z} = \frac{\omega \tau_{max} \delta}{1 + \omega^2 \tau_{max}^2 \delta} \qquad (13)$$

Where $\omega$ is the angular frequency, $\tau_{max}$ is the characteristic time constant that corresponds to the maximum of $\text{Im}Z$ in the Nyquist plot and $\delta$ is the ratio of half sample thickness, d to Debye length, ($\lambda_D$). The maximum of the loss tangent in the $\tan(\varphi) - \log f$ plot appears at,

$$\tan(\varphi)_{max} = \frac{\sqrt{\delta}}{2} \text{ at } \tau_{min} = \tau_{max}\sqrt{\delta} \qquad (14)$$

Where $\tau_{min} = \tau$ and it is the minimum analogous of $\tau_{max}$ or the characteristic time constant at which value $\tan(\varphi)$ attains its maxima in the $\tan(\varphi) - \log f$ plot. The Debye length is then defined from the dimensionless quantity, $\delta$ as

$$\lambda_D = \frac{d}{\delta} = \frac{d}{(2.\tan(\varphi)_{max})^2} \qquad (15)$$

And the diffusion length, $l_D$ is defined as,

$$l_D = \sqrt{\tau_s . D_m} \qquad (16)$$

Where, the time constant, $\tau_s$ corresponds to the maxima of Bode plot ($\text{phase shift}(\varphi) - \log f$) which is also the minima of $\tan(\varphi) - \log f$ plot at low frequency.

## B. Diffusion Coefficient of net mobility and Network structure

AgI-Ag$_2$O-MoO$_3$ glasses are fast ion conductors. Due to applied electric field, cations migrate through the electrolyte and reduce near electrode. The cationic migration model is based on Mott-Gurney (MG) model for electric field driven thermally activated ion hopping and the reduction, charge transfer reaction that leads to electrochemical metallization can be described by Butler-Volmer (BV) equation [32]. Thus, the diffusion model also involves these two processes (i) the inert electrode hinders ionic intercalation between electrode and electrolyte because of high work function but allows electron transfer and instigates the electrode polarization loss processes and double layer formation. The inequality between the size of anodic and cathodic semicircle arcs in the Nyquist plot (**Figure 1B**) suggests a formation of gradient in ion concentration within the system. Hence the system develops a self-diffusion mechanism. The transport model followed in this mechanism is the MG model. This can be represented as $(D_m)_{ionic}$ (ii) Electronic charge transfer reaction between electrode and electrolyte where the transport is based on BV model. This part of diffusion is represented as $(D_m)_{electronic}$. Now, how this diffusion behaves with the structure of the glass?

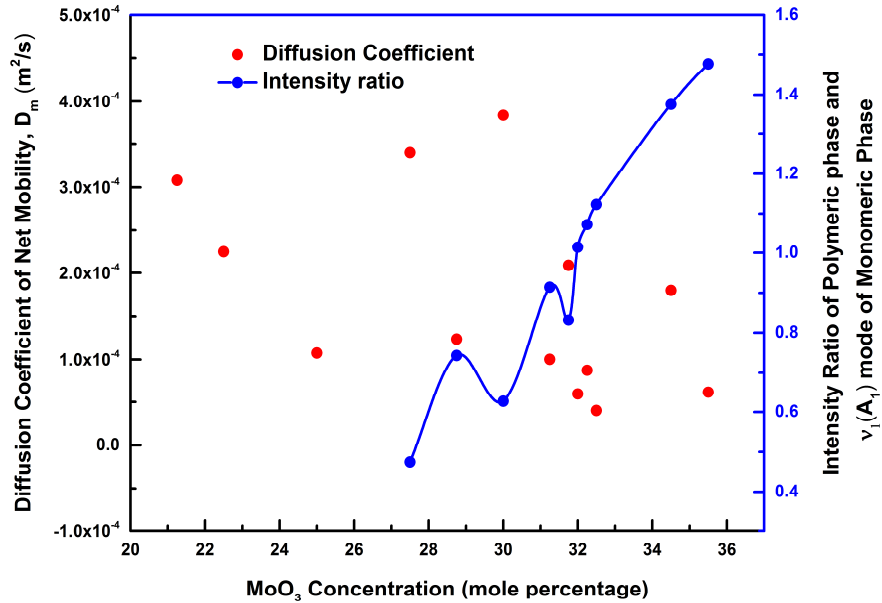

***Figure 4***: *Diffusion coefficient of net mobility, $D_m$ ($m^2/s$) and the Intensity ratio of polymeric phase and symmetric stretching mode ($v_1(A_1)$) of the monomeric phase vs. MoO$_3$ concentration in AgI-Ag$_2$O-MoO$_3$ glass system.*

The Raman spectroscopic data suggests that composition with Ag$_2$O/MoO$_3$ ≥ 1, forms tetrahedral monomeric orthomolybdate ions, [MoO$_4$]$^{2-}$; on the other hand, composition with Ag$_2$O/MoO$_3$ < 1 forms polymeric interconnected chains of linked $MoO_4$ tetrahedra and $MoO_6$ octahedra. The intensity ratio of the obtained polymeric phase and symmetric stretching mode ($v_1(A_1)$) of the monomeric phase provides a measure for the network density or the population of polymeric phase over the monomeric one. **Figure 4** shows the profile of diffusion coefficient and this intensity ratio with respect to the $MoO_3$ concentration. Clearly, the polymeric phase formation doesn't affect the diffusion coefficient. The monotonic increase in the ratio suggests an increase in polymeric phase concentration in expense of monomeric phase (**Equation 1**). On the other hand, $D_m$ exhibits irregular behavior throughout the whole composition range. The scattered nature of $D_m$ is caused by this coupled electronic-ionic contribution. Increasing concentration of $MoO_3$ in the glass matrix causes polymeric phase formation which leads to decrease in non-bridging oxygen (NBO) density. NBO's are the essential hopping sites for the cationic migration. Thus the $(D_m)_{ionic}$ part is affected due to this lacking NBO concentration. This effect of $(D_m)_{ionic}$ on $D_m$ is significantly perturbed by increasing electronic activity. Decreasing NBO and AgI concentration gradually transforms the glass into more electronic in nature. This feature is graphically presented and discussed in later section on conductivity. Thus the total effect glass structure which is composed of this monomeric/polymeric molybdate, remains dissociated from the net diffusion process.

### C. The nature of Relaxation Time Distribution

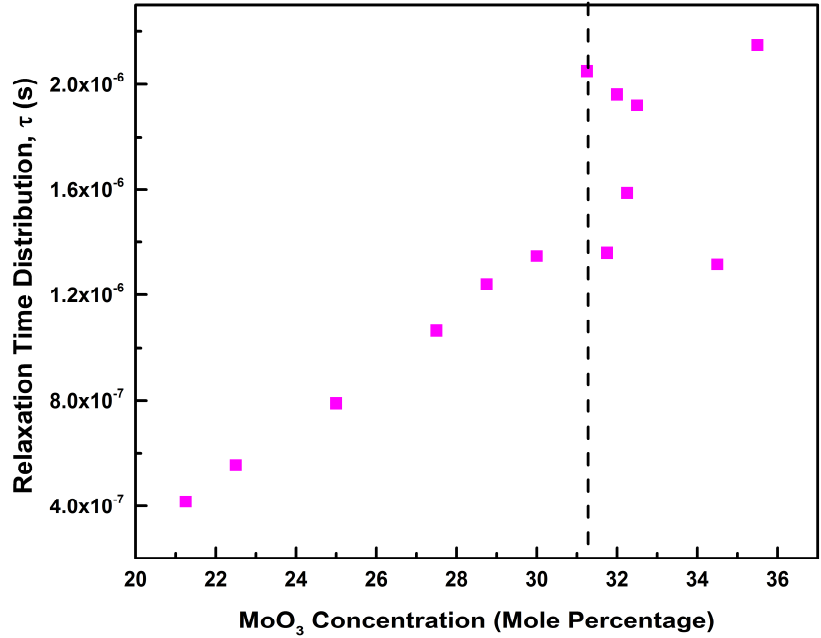

*Figure 5*: *Relaxation time distribution, τ (s) profile over the MoO₃ concentration (Mole percentage)*

Relaxation time, τ is the time that characterizes the motion of a cation between equivalent positions around a NBO [33]; this also infers the time required to rearrange all the induced cation-NBO dipoles along the new orientation. **Figure 5** shows the relaxation time behavior over the composition range. The profile shows that till x ≤ 31.25, τ increases monotonically and reaches a global maximum. Beyond this threshold, the profile gets scattered. Interestingly, this threshold coincides with the fragility threshold obtained in our earlier work [20]; the region below this threshold is comprised of glasses that exhibit nanocluster formation along with a tissue-cluster type structure. As the polymeric phase concentration increases with $MoO_3$ concentration, the glass structure gets over-accumulated by the cluster type structure and eventfully forms a strong glass region.

This behavior of relaxation is explained in the light of Diffusion Controlled Relaxation (DCR) model [17, 18]. High AgI concentration and $Ag_2O/MoO_3 \geq 1$, these two conditions introduce NBOs in a form of vacancies. The inequality between the cation and vacancy concentration determines the hopping probability. When these two conditions are satisfied, this inequality becomes such that it instigates a hopping process from distant sites. This hopping of a distance vagrant ion lessens the τ. While increasing the $MoO_3$ concentration, these vacancies are kept getting filled up. This results in gradual increase in τ until the maximum at the threshold. Beyond this threshold, τ shows scattered nature (mean = 1.715e-6, standard deviation (s.d.) = 3.44e-7). This is caused by a swift change in network characteristics. The threshold determines the margin between two types of glasses: depolymerized fragile region in x ≤ 31.25 (Region I) and polymerized strong region in x > 31.25 (Region II).

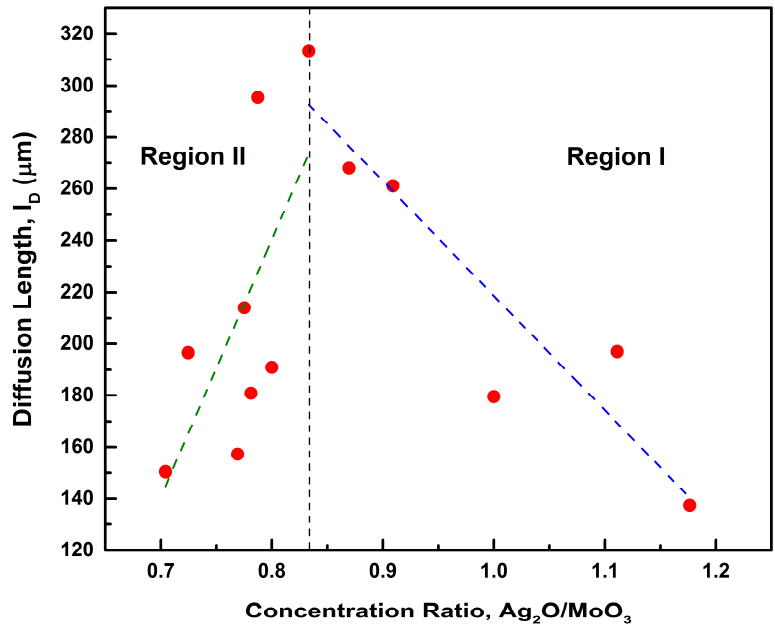

***Figure 6***: *Diffusion length, $l_D$ (μm) vs. the concentration ratio between the alkali oxide ($Ag_2O$) and glass former ($MoO_3$) i.e. $Ag_2O/MoO_3$*

Now, as it is shown in the later section that the A.C. conductivity ($\sigma_{ac}$) exhibits dispersive behavior throughout the selected frequency domain, 40Hz to 10MHz, for all samples, the conductivity process is purely nonrandom, correlated back and forth hopping [34]. Here we assume that the diffusion length ($l_D$) is integral multiple of the length that defines the spatial extent of nonrandom hopping, ($\sqrt{<R^2(t^*)>}$). This quantity can successfully establish the distinction between these two types of glasses: glasses from Region I show faster decrease in $\sqrt{<R^2(t^*)>}$ with increasing alkali ($Ag_2O$) content than glasses from Region II [35]. Hence, **Figure 6**, presenting a plot for $l_D$ vs. $Ag_2O/MoO_3$ shows that Region I has a decreasing tendency in the linear fit. This tendency becomes opposite in Region II. This abrupt change happens when $l_D$ reaches its maximum value at the threshold. Thus, it is noteworthy that Nguyen-Breitkopf method [16] for calculating $l_D$ and $\tau$ has been identified as a novel technique to determine the fragility threshold.

### D. Warburg Coefficient and Loss Resistances

This nature of relaxation reflects upon the behavior of Warburg coefficient, it also shows a similar pattern as $\tau$ **Figure 7(a),** a monotonic increase till $x \leq 31.25$ and a maximum. This suggests that the Warburg coefficient, below this threshold is dominated by the diffusion part, $W_{diff}$ i.e. the effect of this diffusion in the fragile region is much more prominent than in the region beyond. In the $x > 31.25$ region, as the system becomes more electronic in nature due to lack of NBOs, the effect of $W_{ion}$ becomes significant over the diffusion. Similar nature is also observed in the anodic and cathodic loss resistance (**Figure 7(b)**).

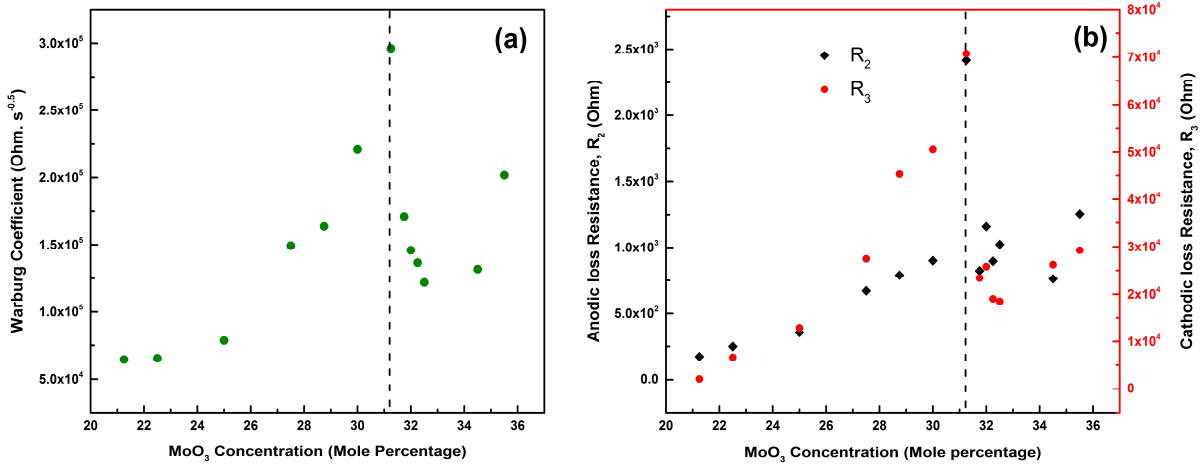

*Figure 7: (a) Warburg coefficient (Ohm. $s^{-0.5}$) vs. $MoO_3$ concentration (Mole percentage) (b) Anodic ($R_2$) and Cathodic ($R_3$) loss resistance vs. $MoO_3$ concentration (Mole percentage)*

### E. Relaxation and $\sigma_{dc}$

The relaxation behavior should reflect directly on D.C. conductivity ($\sigma_{dc}$) because according to DCR model, $\tau$ and $\sigma_{dc}$ are related by a derived form of,

$$\tau = \tau_0 \sigma_0 \exp\left[\frac{2W}{kT}\right] \sigma_{dc} \tag{17}$$

Where, $\tau = \tau_0 \exp\left[\frac{W}{kT}\right]$, W is the activation barrier for diffusion-independent motion [33]. The $\sigma_{dc}$ is calculated from the obtained impedance data for anodic and cathodic loss resistances, $R_2$ and $R_3$, using the relation [36],

$$\sigma_{dc} = \frac{d}{SR_b} \tag{18}$$

Where, d is the sample thickness and S is the surface area. $R_b$ defines the bulk resistance which is in the present situation has been chosen to have a value of ($R_2 + R_3$). **Figure 8** presents the profile of $\sigma_{dc}$ over the whole composition range. It is evident from the behavior of $\sigma_{dc}$ that it has a proportional relation with $\tau$ till the threshold where it reaches its minimum value. Beyond the threshold, $\sigma_{dc}$ exhibits less fluctuating, moderate inclination. The sharp decrease in $\sigma_{dc}$ in the depolymerized region due to increase in $MoO_3$ concentration is due to the filling up of vacancies; the rate of decrement of $\sigma_{dc}$ becomes significantly lessened in the polymerized region. This is analogous to the behavior of $\sqrt{<R^2(t^*)>}$ with alkali oxide concentration.

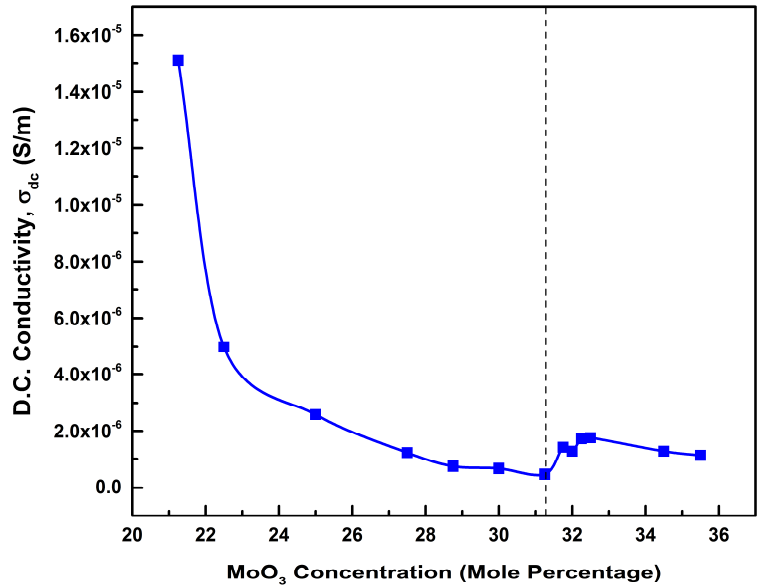

***Figure 8:*** *$\sigma_{dc}$ (S/m) vs. the $MoO_3$ concentration (Mole Percentage)*

**F. Nature of $\sigma_{ac}$ and Power Law behavior**

The A.C. conductivity, $\sigma_{ac}$ is determined from the obtained impedance data for real and imaginary values of impedance, for all samples, at room temperature, by using the relation [36],

$$\sigma_{ac} = \frac{d}{S}\left(\frac{Z'}{(Z'^2 + Z''^2)}\right) \tag{19}$$

**Figure 9** represents the dispersive profile of $\sigma_{ac}$ i.e. $\log(\sigma_{ac})$ with $\log(\omega)$ for a representative sample, $(AgI)_{53.75}$-$(Ag_2O)_{25}$-$(MoO_3)_{21.25}$. This analysis is performed to understand the dispersion relation of $\sigma_{ac}$ as per the universal power law:

$$\sigma_{ac}(\omega) = \sigma(0) + A\omega^n \tag{20}$$

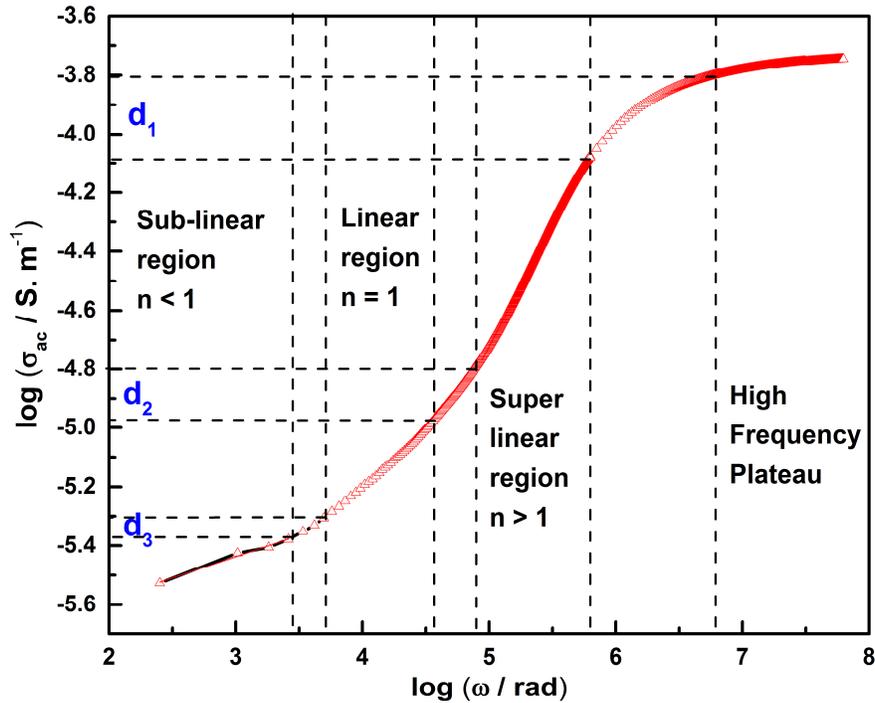

*Figure 9: Dispersion relation of $\sigma_{ac}$; change in $\log(\sigma_{ac}/S.m^{-1})$ with $\log(\omega/radian)$ for a representative sample, $(AgI)_{53.75}$-$(Ag_2O)_{25}$-$(MoO_3)_{21.25}$. $d_1$, $d_2$ and $d_3$ represent the extent of crossover between different dispersive regions.*

Within the impedance scan range, $\sigma_{ac}$ shows diverse frequency dependences.

- Frequency domain I: $2 \leq \log(\omega) \leq 3.5$ exhibits a sub-linear behavior i.e. $n < 1$. This domain is the crossover between the frequency independent D.C. regime and the low frequency dispersive regime. The conductivity in this region is mainly long range ion transport resulting from consecutive hopping. Due to low frequency, polarization effect takes place and enhances the diffusion and hence, $W_{diff}$ part in Warburg.

- Frequency domain II: $3.5 \leq \log(\omega) \leq 4.5$ exhibits a slope of one. This feature of n = 1 is attributed to the fact that the number of mobile ions gets fixed and becomes temperature independent [37].
- Frequency domain III: A super linear power law (SLPL) behavior is observed in the frequency domain $5 \leq \log(\omega) \leq 6$. The appearance of SLPL is striking. SLPL has been associated with a verity of physical processes for different materials [38, 39].

Frequency domain IV: $7 \leq \log(\omega) \leq 8$ shows the high frequency plateau (HFP) region. This region is formed because due to high frequency, the consecutive hop relaxation time would not fit into the time window [36]. Thus each and every hopping contributes to the conductivity [40].

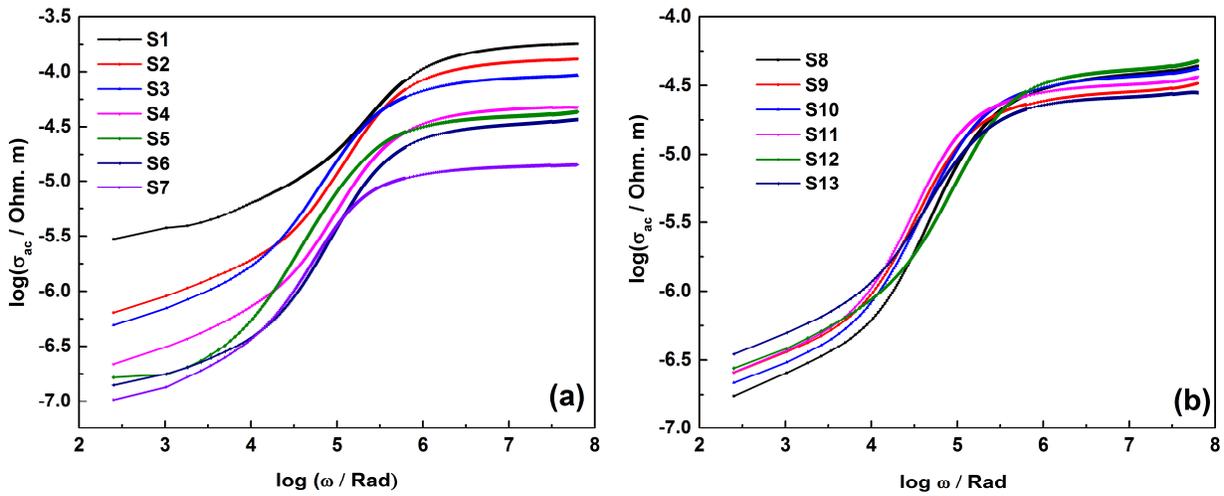

*Figure 10:* $\log(\sigma_{ac} / S.m^{-1})$ vs. $\log(\omega / radian)$ profile for *(a)* samples from Region I *(b)* samples from Region II

**Figure 10** shows the dispersive $\sigma_{ac}$ profile for all the samples, characterized by the nature of network. For all these samples;

- n (mean) = 0.26; s.d. = 0.047 for Frequency Domain I.
- n (mean) = 1.15; s.d. = 0.12 for Frequency Domain III.
- The regular, downward shifting of $\sigma_{ac}$ profiles for Region I samples (**Figure 10(a)**) and grouping for Region II samples (**Figure 10(b)**) is the effect of $\sigma_{dc}$ as described in **equation 20** and Figure 8.

The only significant change between these two regions occurs in the width of Frequency Domain II ($d_{FDII}$), III ($d_{FDIII}$) and the crossover thickness $d_2$. With gradual increase in MoO$_3$ mole percentage, polymerization and lessening NBO concentration, $d_{FDII}$ and $d_2$ shrink and $d_{FDIII}$ broadens. This shrinking nature of $d_{FDII}$ suggests that the number of fixed mobile ions is also decreasing. Hence, it is possible that, in Frequency Domain III, nature of mobile charge carriers changes. This glass neither shows polaronic conduction that should appear in microwave regime, nor lattice mode or electronic excitation that becomes prominent in frequency regime of $> 10^9$ Hz. Thus we propose that the origin of SLPL and broadening of $d_{FDIII}$ is due to new oxygen vacancy formation and their migration. But this proposition requires further research based on other spectroscopic evidences which are beyond the scope of the present work.

5. **Conclusion**

The contributing factors to the diffusion process has been identified to be electronic, which occurs in the electrode-electrolyte interface and ionic, in the bulk. Their mismatch in contribution presents a scattered behavior over the whole composition range, while the intensity ratio of polymeric and monomeric phases exhibits regularity. This happens due to the NBO distribution; the monomeric phase appears in the composition range $Ag_2O/MoO_3 \geq 1$, where higher NBO concentration introduces fragile behavior, significant effect on the spatial extent of nonrandom hopping and instigates vagrant ion hopping. The polymeric phase formation fills up the NBOs and results in $\sigma_{dc}$ and the diffusion process becomes back and forth, correlated hopping. This modification in diffusion behavior results in Warburg and electrode loss resistances. Moreover, the power law behavior of $\sigma_{ac}$ for all samples shows non-Jonscher type behavior with the sub-linear exponent 0.26, super-linear exponent 1.15, a slope 1 crossover and a high frequency plateau. $\sigma_{ac}$ profile lacks behavioral changes over all compositions other than shrinking of slope 1 regime and broadening of SLPL; this phenomenon has been intuitively corroborated with the oxygen vacancy formation at high frequency.

- **Acknowledgements**

We thank facility technologists Ashwini Kumari AS for Impedance Spectroscopy and Pradeep Kumar M. L. for Raman spectroscopy (MNCF, CeNSE, IISC Bangalore). We convey our thanks to Dr. Tathagata Biswas (Arizona State University, IISc Bangalore (Former)) for helping us to use the supercomputer platform.